\documentclass[a4paper]{elsarticle_rfabbri}
\pdfoutput=1
\usepackage{amssymb,amsmath,verbatim}
\usepackage[english]{babel}    
\usepackage{subfigure}
\usepackage{booktabs}
\usepackage{picinpar}
\usepackage{color,xcolor}
\usepackage{lineno}
\usepackage{float}

\newcommand{\ie}{{\it i.e.}}

\newcommand{\eg}{{\it e.g.}}

\newcommand{\etal}{{\it et.\ al.\ }}

\usepackage[pagebackref=true,breaklinks=true,letterpaper=true,colorlinks,bookmarks=false]{hyperref}

\journal{ }

\begin{document}

\begin{frontmatter}

\title{Multi-$q$ Pattern Classification of Polarization Curves}

\author[iprj]{RICARDO FABBRI}
\ead{rfabbri@iprj.uerj.br}

\author[iprj]{IVAN N.\ BASTOS}
\ead{inbastos@iprj.uerj.br}

\author[iprj]{FRANCISCO D. MOURA NETO}
\ead{fmoura@iprj.uerj.br}

\author[ufrj]{FRANCISCO J. P. LOPES}
\ead{flopes@ufrj.br}

\author[ifsc]{WESLEY N. GON\c{C}ALVES}
\ead{wnunes@ursa.ifsc.usp.br}

\author[ifsc]{ODEMIR M. BRUNO}
\ead{bruno@ifsc.usp.br}

\address[iprj]{Instituto Polit\'{e}cnico, Universidade do Estado do Rio de Janeiro\\C.P.: 97282 - 28601-970 - Nova Friburgo, RJ, Brazil}

\address[ufrj]{Instituto de Biof\'{i}sica Carlos Chagas Filho, Universidade Federal do Rio de Janeiro\\
Sala G1-019 - Cidade Universit\'{a}ria, 21941-902 - Rio de Janeiro, RJ, Brazil}

\address[ifsc]{Instituto de F\'{i}sica de S\~{a}o Carlos (IFSC), Universidade de
S\~{a}o Paulo (USP)\\ Av.  Trabalhador S\~{a}o Carlense, 400,
13560-970 - S\~{a}o Carlos, SP, Brazil}

\begin{abstract}
Several experimental measurements are expressed in the form of one-dimensio-nal
profiles, for which there is a scarcity of methodologies able to classify the
pertinence of a given result to a specific group.  The polarization curves that
evaluate the corrosion kinetics of electrodes in corrosive media are an
application where the behavior is chiefly analyzed from profiles.  Polarization curves
are indeed a classic method to determine the global kinetics of metallic
electrodes, but the strong nonlinearity from different metals and alloys can
overlap and the discrimination becomes a challenging problem. Moreover, even
finding a typical curve from replicated tests requires subjective judgement. In
this paper we used the so-called multi-$q$ approach based on the Tsallis
statistics in a classification engine to separate multiple polarization curve
profiles of two stainless steels.  We collected 48 experimental polarization
curves in aqueous chloride medium of two stainless steel types, with different
resistance against localized corrosion.  Multi-$q$ pattern analysis was then
carried out on a wide potential range, from cathodic up to anodic regions.
An excellent classification rate was obtained, at a success rate of $90\%$,
$80\%$, and $83\%$ for low (cathodic), high (anodic), and both potential ranges, respectively,
using only $2\%$ of the original profile data.
These results show the potential of the proposed approach towards
efficient, robust, systematic and automatic classification of highly non-linear
profile curves.
\end{abstract}


\end{frontmatter}

\section{Introduction}
Several results of experimental techniques in materials science and engineering are available in
the form of profiles. Despite the crucial importance of these types of data, the use of
statistics on profiles is rare, as the adequate statistical techniques
themselves are under
development~\cite{Zhang:et:Albin:2009,Magalhaes:etal:2009,Magalhaes:et:MouraNeto:2012}. Useful applications of profile analysis
include outlier detection and classification/clustering of groups of results,
taking into account the inherent stochasticity of experimental data.

Since the use of statistics for realistic profile datasets is currently somewhat
scarce, the evaluation and comparison of experimental results is ordinarily
performed by subjective criteria, \eg, after performing three to five runs of an
experiment to identify representatives.  In the case of profiles, a physical
model with a related adjustment of parameters is a good indication of pertinence
to a target group. In this situation, however, the analysis is mainly focused on the
parameter values instead of profiles themselves. We here explore systematic data
analysis techniques suited for \emph{complex} or \emph{highly nonlinear}
profiles, \ie, when there is no simple model description or parameters to
represent the entire profile curve.

In the fields of corrosion and materials science, there are numerous examples of data 
with complex profile shapes, for instance: the aforementioned polarization curves, impedance diagrams,
voltamograms, electrochemical noise records and so on. Despite this profusion of
profile data, the work of Strutt~\etal~\cite{Strut:etal:CS1985} is a rare example of profile
analysis from corrosion, albeit the studied profiles were related to the geometric shape of
corroded interfaces, and not electrochemical data as in the present work.

Polarization curves, or, more generally, current-potential plots, have
crucial importance in corrosion studies as well as in electrochemistry. They are
essential for measuring the global kinetics of electrodes. A number of
electrochemical parameters can be obtained from these tests, from charge transfer,
passivation, corrosion current density, to mass transport properties. As
expected, for any set of experimental results, dispersion always occurs.  Even a
stable reaction performed under stringent control of experimental procedures,
such as the ferri/ferrocyanide redox reaction, can present significant
scattering~\cite{Tourwe:Breugelmans:etal:2007}. Moreover, the level of
scattering in polarization curves naturally increases as the localized corrosion
takes place, which exhibits random nature. Another challenging case is that the surface of commercial grades
of steels, that has wide industrial interest, has inherent heterogeneity of their
metallurgical aspects. Furthermore, certain stochastic phenomena such as pitting
produce electrochemical current variations that produce overlap in the experimental
curves, rendering them almost inseparable.  However, these flutuactions can also
be very useful to describe the type and the intensity of the corrosion
attack~\cite{Bastos:Nogueira:MCP2008,Klapper:etal:CorrosionScience2010}. Thus,
for some corrosion resistant alloys (CRA) such as stainless steels, the
evaluation of parameters from polarizations curves near pitting potential is
an important procedure to classify their corrosion resistance.

Having had success in employing Tsallis statistics~\cite{Tsallis:book:2009} to
the classification of image patterns as well as other
applications~\cite{Fabbri:etal:PhysicaA2012}, we
devise a similar multi-$q$ approach based on such statistics to propose a method to
classify highly nonlinear profiles from corrosion tests. In the present work we classify 48
polarization curves produced from two stainless steel classes, with different
resistance against localized corrosion, in aqueous chloride medium,
Figure~\ref{fig:polarization:curves}.  The primary goal of
any classification method is to assign a class label (\eg, material type) to a given data sample (\eg, image or
profile). In our case, the profiles are the polarization curves and the output
label is one of two stainless steels. Despite having slightly different chemical
compositions, the polarization of both materials is very similar, especially in
the cathodic region and in the potential immediately past the corrosion potential region. This
similarity produces overlapping curves which require a systematic approach to
classify, an analysis that has not been routinely carried out in this field.

In statistical mechanics, the concept of entropy is used to characterize a
system based on its distribution of states, which can be determined by the
system energy level. If one has less information about the system then at many
states the system can still be characterized. This idea produces a direct link
between the concept of entropy and the lack of information of a system. From an
information-theoretical point of view, the entropy represents how close a given
probability distribution is to the uniform distribution. Since the
kinectic aspects of
polarization curves are related to allowed 
microstates of the system, an approach based on the concept of entropy
is a natural choice. In addition, the random nature of this kind of
experimental data is caused, among other factors, by the joint diversity of the
microstates, which is directly related to the different chemical composition of
the materials and the superficial conditions, chiefly near the pitting
potential.  Because such chemical compositions are expected to produce
both short- and long-range interaction in the
material~\cite{Punckt:etal:Science2004}, using the Tsallis entropy, which
introduces a new parameter $q$ to help model these interactions across different
scales, motivated our approach.

Formally, the Tsallis entropy for a given profile curve and parameter $q$ is defined as
\begin{equation}
S_{q} = \frac{1 - \sum p_i^q}{q-1},
\end{equation}
where ${p_i}$, $i=1,\dots,N$ is an estimated probability distribution of desired pattern
properties of a signal. In this paper we use the histogram of measurement
values in the quantized profile to represent pattern properties, where the
profile amplitude was quantized in $N$ levels of the same size.
The multi-$q$ analysis consists in expressing statistical aspects of this
histogram as a vector of entropy values for $n < N$ different values of $q$,
\eg, $0.1,0.2,\dots,q_{n}$. This is called the multi-$q$ vector of the
profile, which is much like an entropic spectrum. In practice, best-performing
values for $n$ are typically very small (\eg, 10), two to three orders of
magnitude less than the number of raw data (profile) sample points (\eg, 800 in the
present paper).

\section{Materials and Methods}

\subsection{Setup for Generating Polarization Curve Data}
Two commercial austenitic stainless steels (UNS
S30400 and UNS S31600) were used, hereafter named 304 and 316 for simplicity.
 Bar samples were coated with $\text{Teflon}^{TM}$ and an
area of $0.20 cm^2$ was kept accessible to the electrolyte. Before each test, the
samples were grit with emery paper up to $\#600$, and subsequently washed with
distilled water and dried with hot air. These two stainless steels have
austenitic microstructure, but their main difference being the molybdenum
content. The 316 alloy has \emph{circa} 2.5\% Mo that improves its pitting resistance
in chloride media. The 304 has only a residual content of molydbenum. 
The aerated aqueous electrolyte was 3.5\% in mass of NaCl with
temperature kept within $25.0\pm0.1^\circ C$. 

All experiments were performed inside a Faraday cage in order to avoid spurious
electric noise. After the immersion of the sample in the solution, the assumed
steady-state condition is attained after an hour at open-circuit status. After this step,
potentiodynamic polarization was applied from $-0.6$ up to $0.2\, V \times SCE$
under a potential rate of $1.0\, mV s^{-1}$. All values of the potential were measured with a
saturated calomel electrode (SCE) reference. A platinum wire was used as
counter-electrode. In order to obtain sufficient data, 24 curves were performed
for each alloy.  Each polarization curve consists of a profile of potential
versus current (or, equivalently for a constant area, current density) as a
sequence of 800 data points.

\subsection{Classification Approach}

We performed profile pattern classification experiments whose basic goal is to
assign a steel class label to any given profile curve for which
the steel type is previously unknown. The classification approach is data-driven and
supervised, in that the model is not specific for stainless steel polarization
curves, but instead is learned and adjusted/trained from previously seen
data~\cite{Fabbri:etal:PhysicaA2012}.  In supervised classification, the
classifier is trained from a set of samples that are known to belong to the
classes (\emph{a priori} knowledge)~\cite{Duda:etal:2001}. The classifier is
then validated by a different set of so-called `test' samples for which we know
the ground-truth steel classes but have not been used to train the classifier.
For each test sample the classifier outputs a likely steel type and we generate
a success rate based on how well these match to ground-truth.

We employed a number of minor variants of our so-called multi-$q$ classification
approach to classify the signals, which consists in using Tsallis
statistics~\cite{Fabbri:etal:PhysicaA2012} on the profile data followed by the
naive Bayes classifier~\cite{Duda:etal:2001}. We have also used classical
principal component analysis (PCA)~\cite{Costa:Cesar:Book2010} as an optional
step. PCA consists in expressing each original profile (800 points) in a rotated
and translated coordinate frame such that the entire dataset variance along each
direction is maximum. The desired dimension may be reduced to less than 800
by keeping only the coordinates of highest variance. The goal of this
reduction is threefold: representational efficiency, visualization purposes, and
for automatically filtering out low-variance aspects of data that may otherwise
add up to hinder classification performance.

In order to objectively evaluate the performance of the classification approach,
we use the stratified 10-fold cross-validation scheme~\cite{witten2011data}.  In
this scheme, the profile dataset is randomly divided into 10 folds, considering that
each fold contains two profiles, one for each steel class.  At each run
of this scheme, the classifier is trained using all but one fold and then is
evaluated on how it classifies the samples from the separated fold.  This
process is repeated such that each fold is used once as validation.  The
performance is averaged, generating a single number for classification rate
which represents the overall proportion of success over all runs.  

\section{Results and Discussion}
\begin{figure}[ht]
\centering
\includegraphics[width=1\textwidth]{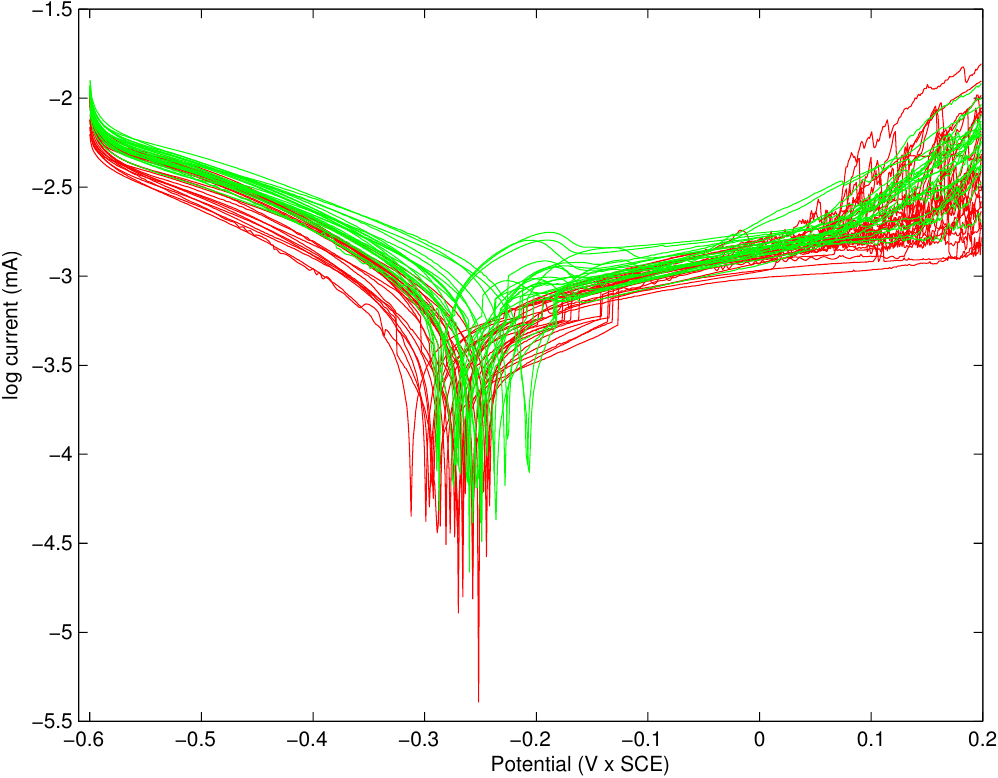}%
\caption{%
48 experimental polarization curves of stainless steels 304 (red) and 316 (green). Each curve
consists of 800 sample points. 
}\label{fig:polarization:curves}
\end{figure}

Figure~\ref{fig:polarization:curves} shows the 48 polarization curves of stainless steels 304 and 316 swept from cathodic to anodic regions.  At actual corrosion situation, cathodic and anodic processes occur on the same metallic surface, under equal intensity. Thus, to obtain a global view of the electrode kinect we normally apply a potential range that covers both regions. Under cathodic potential, there is a predominance of cathodic process on work-electrode, that is a fixation of electrons, which in the present case is the reduction of dissolved oxygen gas. On the other hand, under anodic polarization, the net current provokes the corrosion of alloy, an oxidation process. At low anodic potential the corrosion process tends to be uniform, and at high potential it can present a localized corrosion such as pitting that has essentially a stochastic nature. Between these regions, the net current approachs null value, thus the logarithm of current attains  the lowest level, defining the corrosion potential. The cathodic process occurs to the left and the anodic to the right side of the corrosion potential.
The polarization of two stainless steels in
aerated solution bears certain similarities, at least in the cathodic region and for potentials just
beyond the corrosion potential. Therefore, an overlapping of curves can be observed,
despite their different chemical composition. However, close to potential
where there is a higher susceptibility to pitting, the difference of resistance
appears together with some natural dispersion. 
Nevertheless, as in Figure~\ref{fig:polarization:curves}, we can see a significant overlap,
 when a large number of curves is considered.
Therefore, at first sight, one cannot easily assign a given curve to the
related steel type. Moreover, 304 exhibits a higher scattering of current above
\emph{circa} 
$0.0  V \times SCE$. In that region, a highly stochastic event occurs: the
pitting attack. The 316 steel type is specially designed to resist this localized
corrosion.

Mild evidence of the high resistance of 316 can be observed at a corrosion
potentials, with mean value \emph{circa} $-250 mV \times SCE$, being slightly higher for 316 than 304, which fares close to $-280 mV \times SCE$.  However,
the main stochastic phenomenon occurs for higher potentials, where the randomness aspect is strong
due to pitting. The main interest in stainless steel corrosion behavior is limited to this high
potential range. In this case, for a small number of experiment runs, as in ordinary
practice, it is possible to classify erroneously one steel instead of the other,
as the most pitting resistant,
in a false-positive or similarly false-negative evaluation. Thus, judgement
bias can take place due to the small number of observations. There is no
generally accepted theory on how exactly these stochastic
flunctuations occur~\cite{Balazs:Gouyet:PhysicaA1995}, despite their huge
industrial importance. To properly study this fact,
the advisable number of test runs is far beyond the usual standard of
subjective procedures which rely on only 3-4 runs.  

In addition to the stochastic events in the region of interest (pitting), in this case the
current is the result of interest, and higher currents mean stronger corrosion. At
high potential, \emph{circa} $0.2 V \times SCE$, the so-called random localized
attack occurs, and the possibility of wrong interpretation is further aggravated.
Our objective in this
paper is to develop and test a method to classify very nonlinear profiles such
as these, a similar approach also being useful to perform outlier detection.
Thus, the obtained polarization curves of stainless steels in chloride media are
completely adequate to the purposes of our investigation. 

\begin{figure}[t]
\centering
\includegraphics[width=0.8\textwidth]{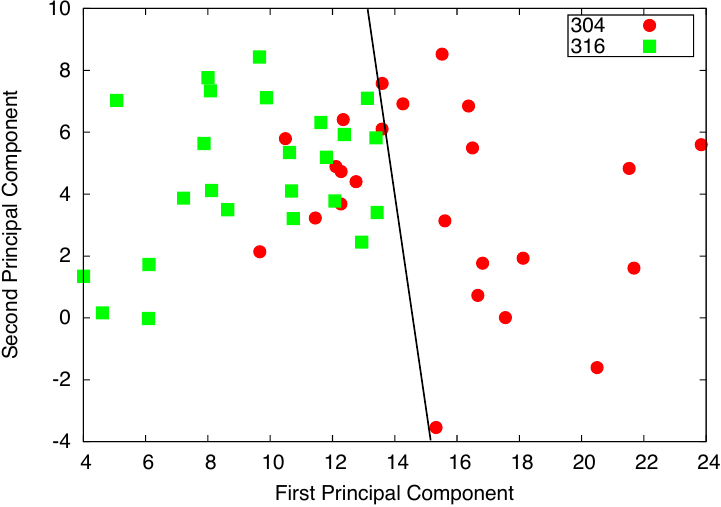}%
\caption{%
\textbf{Results from a direct classical approach, full potential range.} First and second principal
components for a classic principal component analysis on raw profile data,
together with a plausible classification boundary.  
}\label{fig:ivan2}
\end{figure}

Figure~\ref{fig:ivan2} depicts the first and second principal component of the
obtained polarization curves by using classical principal component analysis
directly on raw profile data. A
reasonable boundary separating the two sets of polarization curves was obtained, 
with, however, some degree of
misclassification which may be aggravated if a larger number of experiments where available.
Moreover, 316 shows
a smaller variance (closer principal components) than 304. The relatively small
cluster extent of 316 suggests less susceptibility to stochastic corrosion of
pits, which is consistent with the literature~\cite{Strut:etal:CS1985}. A
plausible classification boundary is drawn with the sole purpose of guiding the
eye to the clusters.
\begin{figure}[H]
\centering
\includegraphics[width=0.8\textwidth]{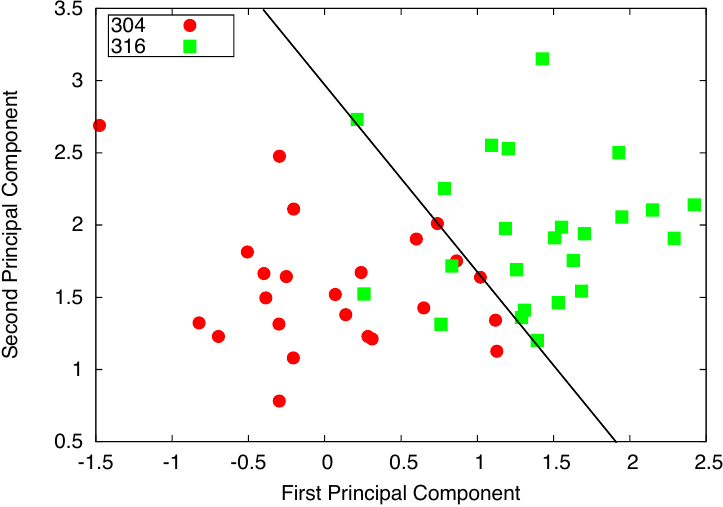}%
\caption{%
\textbf{Results from the proposed approach, full potential range.} The first and second principal
components for the proposed Tsallis entropy approach with $q$ ranging from 0.1
to 2.0 in steps of 0.1, together with a plausible classification boundary. Even with 2
dimensions per profile for ease of visualization, only three 316 curves (green squares) 
are confused with the 304 ones (red circles).
}\label{fig:ivan3}
\end{figure}

As an attempt at improving this separation, we performed so-called multi-$q$
pattern analysis, as shown in Figure~\ref{fig:ivan3} for only two principal
components for visualization purposes (\ie, using two classification dimensions
per profile). In this approach, with $q$ varying from 0.1 to 1.0, a very good
separation was found. 

Table~\ref{tab:ivan1} depicts a more comprehensive summary on the score of several
classification tests. The Tsallis entropy approach, even for as few as two
principal components, enabled classifying most polarization curves
correctly. Moreover, the profile data can be efficiently represented by at most
20 Tsallis entropy values, as opposed to all 800 points. As the table shows, it
was possible to attain the best classification rate using only a single value of $q =
1$. However, using the multi-$q$ vector (bold line) the optimal selection of this parameter
is fully automatic for a given dataset with no observed decrease in classification rate.

\begin{table}
  \renewcommand{\arraystretch}{1.25}
	\caption{Classification rates for a number of different methods with variable
  $q$ and also classifying directly over raw data. The bold line highlights our best
  proposed result, which differs from the first line in that it is
  automatic and less dependant on the application domain, as explained in the
  text.}
  \vspace{1em}
	\centering
  \footnotesize
		\begin{tabular}{c c c c}
			\toprule
			\textbf{Method} & \multicolumn{3}{c}{\textbf{Classification Rate (\%)}}
        \\
      & \textbf {Full Potential} & \textbf {Low Potential} & \textbf {High
      Potential} \\\midrule
			Tsallis $q = 1$ & 83 & 83 & 65 \\
			Tsallis $q = 0.1$ & 73 & 65 & 60\\
			Multi-$q$, $q=0.1,0.2,\dots,1.0$ & 73 & 69 & 69 \\
			\textbf{Multi-$\boldsymbol q$}, $\boldsymbol{q=0.1,0.2,\dots,2.0}$&
      \textbf{83} & \textbf{90} & \textbf{80}\\
			Na\"ive Bayes on all 800 points & 81  & 75 & 73\\
			\bottomrule
		\end{tabular}
	\label{tab:ivan1}
\end{table}

\subsection{Classifying Low and High Potential Ranges Separately} 
\begin{figure}[htb]
\centering
\includegraphics[width=0.8\textwidth]{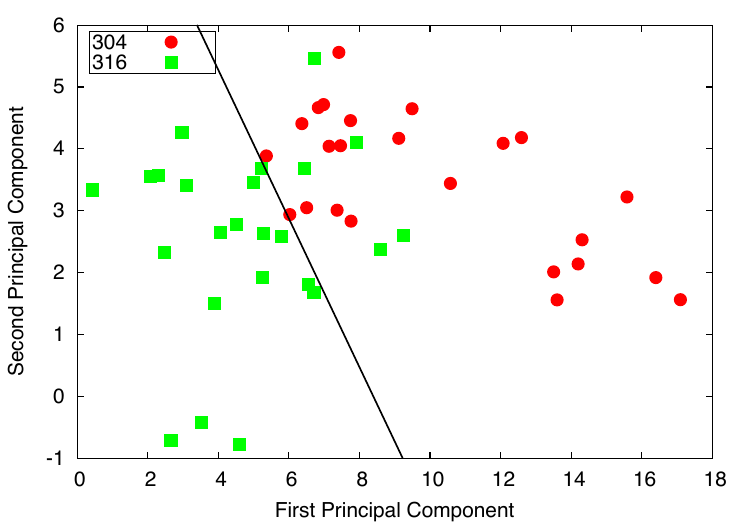}%
\caption{%
\textbf{Results from a direct classical approach, low potential range.} First and second principal
components for a classic principal component analysis on raw profile data,
together with a plausible classification boundary.  
}\label{fig:ivan2:low}
\end{figure}

\begin{figure}[htb!]
\centering
\includegraphics[width=0.8\textwidth]{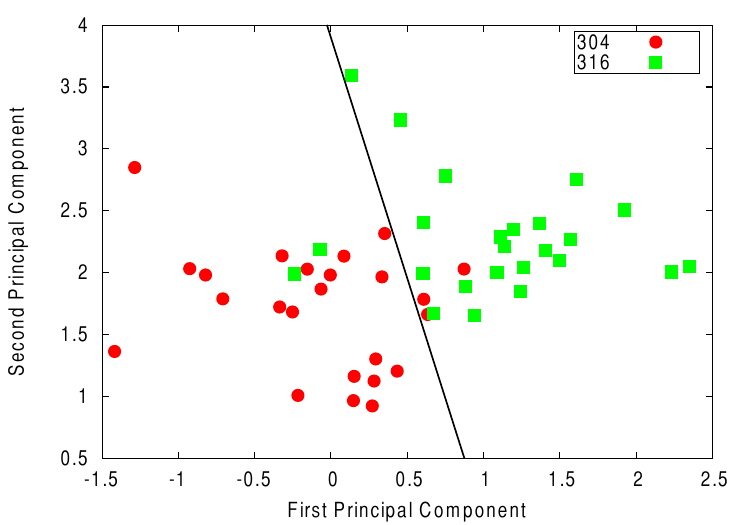}%
\caption{%
\textbf{Results from the proposed approach, low potential range.} The first and second principal
components for the proposed Tsallis entropy approach with $q$ ranging from 0.1
to 2.0 in steps of 0.1, together with a plausible classification boundary. 
}\label{fig:ivan3:low}
\end{figure}

We also performed classification experiments to observe the behavior of the
proposed approach on different sections of
the profile which have distinct statistical properties. 
Classification results using only the first half of the profiles, with
potentials having less than $-0.2 V\times SCE$, are shown
in Figures~\ref{fig:ivan2:low} and~\ref{fig:ivan3:low}. These potentials are
roughly related to the predominance of cathodic ($-0.6\text{ to}\, -0.2 V\times SCE$) and
anodic ($-0.2\text{ to } 0.2 V\times SCE$) processes on both steels.

\begin{figure}[htb]
\centering
\includegraphics[width=0.8\textwidth]{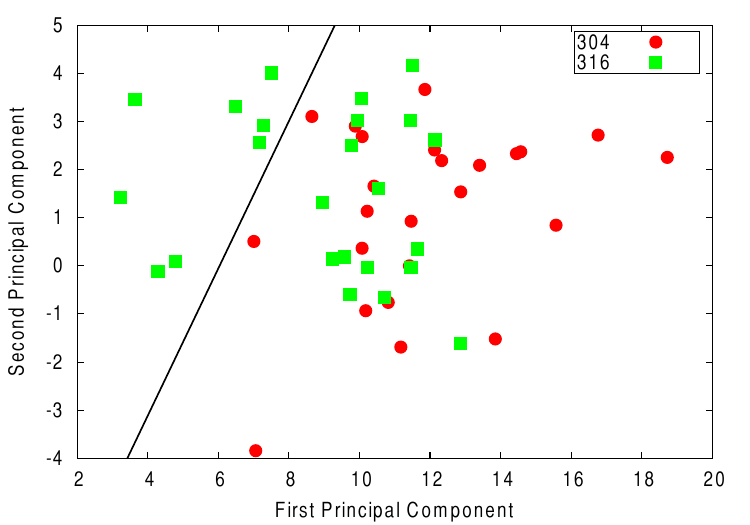}%
\caption{%
\textbf{Results from a direct classical approach, high potential range.} First and second principal
components for a classic principal component analysis on raw profile data,
together with a plausible classification boundary.  
}\label{fig:ivan2:high}
\end{figure}

\begin{figure}[htb!]
\centering
\includegraphics[width=0.8\textwidth]{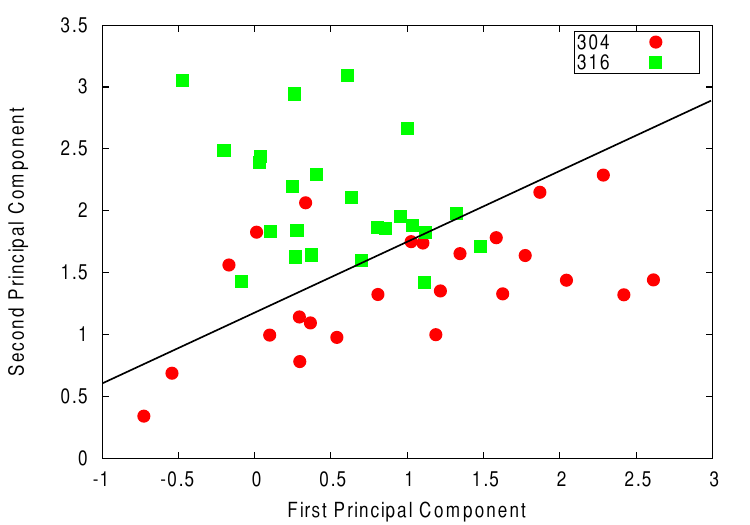}%
\caption{%
\textbf{Results from the proposed approach, high potential range.} The first and second principal
components for the proposed Tsallis entropy approach with $q$ ranging from 0.1
to 2.0 in steps of 0.1, together with a plausible classification boundary. 
}\label{fig:ivan3:high}
\end{figure}

Similar results for
potentials above $-0.2 V\times SCE$ are shown in Figures~\ref{fig:ivan2:high}
and~\ref{fig:ivan3:high}. More comprehensive results are shown in the last two
columns of
Table~\ref{tab:ivan1}. For both low and high potentials, the proposed method
achieved very good
classification rates, having an edge of $7\%$ to $15\%$ over the best success rate
without using Tsallis statistics. The results also indicate that the profiles
are significantly more challenging to distinguish in the high potential range,
which is consistent with the theory and can be confirmed by visual inspection of
Figure~\ref{fig:polarization:curves}, just where its behavior is more meaningful
for technological use.

\subsection{Further Discussion} 
With the present methodology, the classification of strongly nonlinear curves
with a systematic procedure is made possible. The success in classifying a profile
that originally contains 800 points using only a couple or so dimensions could be
understood as a signal expressed in a more suitable statistical representation and
compressed and filtered with a low
sampling frequency. In this case, a natural question is the possibility of
aliasing~\cite{Bastos:etal:JES2000} in the data. However, the first and second
principal components are not subsampled point from the set, but parameters
derived from the statistics, specifically the multi-$q$ vector of Tsallis
entropies, optionally followed by the projection of maximum variance given by PCA.

It is clear that the best classification is attained with a wider multi-$q$
vector with $q$ ranging  from 0.1 to 2.0, instead of a fixed $q$. Moreover, even
though the corrosion behavior at high potential is of great interest, the best
classification rate is obtained with the entire prolarization curves, and not
only at high potentials (anodic process). In this case, from
Table~\ref{tab:ivan1}, the values are 83\% for the entire curve against 80\%
for the anodic half of the potential. Thus, for the purpose of a reliable
classification, polarization curves performed from cathodic to anodic potentials
convey more information than any single range.

\section{Conclusion}
A relatively high number of polarization curves were generated through
experiments performed for austenitic
stainless steels UNS S30400 and UNS S31600 in chloride medium at $25^\circ C$.
For the entire potential range of the polarization curves a degree of overlap of experimental curves was
observed. The data samples were subsequently tested with a Tsallis-based
statistics in order to classify the profiles for each steel. For the case where a
multi-$q$ pattern analysis was carried out on a complete potential range, from
cathodic up to anodic regions, an excellent classification rate was obtained,
using only $2\%$ of the original profile data.  When automatically classifying
only the low potential range, the approach attained $90\%$ correct
classification, as opposed to $80\%$ at
high potentials, which is consistent with the theory of pitting distribution. However, the industrial
interest for these steels lies on the high potential range. These results together
with the inspection of a series of scatter plots using principal component analysis demonstrate
the capacity of the proposed approach towards efficient, robust, systematic and
automatic classification of highly non-linear profile
curves, which can be used for analyzing pattern properties as well as for
outlier detection.

With the promising results obtained in the classification of polarization
curves, an extension to study further types of signals such as time records of
electrochemical noise is ongoing in our research group. In these aspects, the
Tsallis statistics can be very useful to identify the corrosion status from a
continuous recording of on-line monitoring. In this scenario, noisy potential
and current related to corrosion process would be recorded in order to detect changes
that affect the corrosion attack.  

\section*{Acknowledgements}
\small{The authors thank Mr. Denisar Ism\'{a}rio for the electrochemical tests. We
acknowledge the financial support of the Brazilian agencies CNPq, FAPERJ and
FAPESP.  R.F.\ acknowledges support from FAPERJ/Brazil 111.852/2012,
W.N.G.\ from FAPESP (2010/08614-0),
O.M.B.\ from CNPq (Grant \#308449/2010-0 and \#473893/2010-0) and FAPESP (Grant
\# 2011/01523-1).}


\begin{thebibliography}{10}
\expandafter\ifx\csname url\endcsname\relax
  \def\url#1{\texttt{#1}}\fi
\expandafter\ifx\csname urlprefix\endcsname\relax\def\urlprefix{URL }\fi
\expandafter\ifx\csname href\endcsname\relax
  \def\href#1#2{#2} \def\path#1{#1}\fi

\bibitem{Zhang:et:Albin:2009}
H.~Zhang, S.~Albin, Detecting outliers in complex profiles using a $\chi^2$
  control chart method, IIE Transactions 41 (2009) 335¿--345.

\bibitem{Magalhaes:etal:2009}
M.~De~Magalh\~{a}es, A.~Costa, F.~Moura~Neto, A hierarchy of adaptive control
  charts, International Journal of Production Economics 119~(2) (2009)
  271--283.

\bibitem{Magalhaes:et:MouraNeto:2012}
M.~De~Magalh\~{a}es, F.~Moura~Neto, A {L}aplacian spectral method in phase {I}
  analysis of profiles, Applied Stochastic Models in Business and Industry 28
  (2012) 251--263.

\bibitem{Strut:etal:CS1985}
J.~Strutt, J.~Nicholls, B.~Barbier, The prediction of corrosion by statistical
  analysis of corrosion profiles, Corrosion Science 25~(5) (1985) 305--315.

\bibitem{Tourwe:Breugelmans:etal:2007}
E.~Tourw{\'e}, T.~Breugelmans, R.~Pintelon, A.~Hubin, Extraction of a
  quantitative reaction mechanism from linear sweep voltammograms obtained on a
  rotating disk electrode. part ii: Application to the redoxcouple, Journal of
  Electroanalytical Chemistry 609~(1) (2007) 1--7.

\bibitem{Bastos:Nogueira:MCP2008}
I.~N. Bastos, R.~P. Nogueira, Electrochemical noise characterization of
  heat-treated superduplex stainless steel, Materials Chemistry and Physics
  112~(2) (2008) 645--650.

\bibitem{Klapper:etal:CorrosionScience2010}
H.~S. Klapper, J.~Goellner, A.~Heyn, The influence of the cathodic process on
  the interpretation of electrochemical noise signals arising from pitting
  corrosion of stainless steels, Corrosion Science 52~(4) (2010) 1362--1372.

\bibitem{Tsallis:book:2009}
C.~Tsallis,
  \href{http://books.google.com.br/books?id=K7xOhNeGk6kC}{Introduction to
  Nonextensive Statistical Mechanics: Approaching a Complex World}, Springer,
  2009.
\newline\urlprefix\url{http://books.google.com.br/books?id=K7xOhNeGk6kC}

\bibitem{Fabbri:etal:PhysicaA2012}
R.~Fabbri, W.~Gon{\c{c}}alves, F.~Lopes, O.~Bruno, Multi-$q$ pattern analysis:
  A case study in image classification, Physica A: Statistical Mechanics and
  its Applications 391~(19) (2012) 4487--4496.
\newblock \href {http://dx.doi.org/10.1016/j.physa.2012.05.001}
  {\path{doi:10.1016/j.physa.2012.05.001}}.

\bibitem{Punckt:etal:Science2004}
C.~Punckt, M.~B{\"o}lscher, H.~Rotermund, A.~Mikhailov, L.~Organ, N.~Budiansky,
  J.~Scully, J.~Hudson, Sudden onset of pitting corrosion on stainless steel as
  a critical phenomenon, Science 305~(5687) (2004) 1133--1136.

\bibitem{Duda:etal:2001}
R.~Duda, P.~Hart, D.~Stork, Pattern classification, Vol.~2, Wiley New York,
  2001.

\bibitem{Costa:Cesar:Book2010}
L.~da~Fontoura~Costa, R.~M. Cesar~Jr, Shape analysis and classification: theory
  and practice, CRC press, 2010.

\bibitem{witten2011data}
I.~Witten, E.~Frank, M.~Hall, Data mining: Practical machine learning tools and
  techniques, 3rd Edition, Morgan Kaufmann, 2011.

\bibitem{Balazs:Gouyet:PhysicaA1995}
L.~Balazs, J.~Gouyet, Two-dimensional pitting corrosion of aluminium thin
  layers, Physica A: Statistical Mechanics and its Applications 217~(3) (1995)
  319--338.

\bibitem{Bastos:etal:JES2000}
I.~Bastos, F.~Huet, R.~Nogueira, P.~Rousseau, Influence of aliasing in time and
  frequency electrochemical noise measurements, Journal of The Electrochemical
  Society 147~(2) (2000) 671--677.

\end{thebibliography}

\vspace{3em}

\end{document}